\documentclass[a4paper]{aa}
\usepackage{txfonts,graphics,natbib}
\bibpunct{(}{)}{;}{a}{}{,}
\newcommand{\centotrentatre}{133P/Elst-Pizarro}

\newcommand{\pq}{\ensuremath{P_Q}}
\newcommand{\pu}{\ensuremath{P_U}}

\newcommand{\pn}{\ensuremath{P^\mathrm{N}_\mathrm{Q}}}
\newcommand{\pcoma}{\ensuremath{P^\mathrm{C+T}_\mathrm{Q}}}
\newcommand{\pobs}{\ensuremath{P^\mathrm{\,obs}_\mathrm{Q}}}

\newcommand{\Ntot}{\ensuremath{\mathcal{N}_\mathrm{tot}}}
\newcommand{\NN}{\ensuremath{\mathcal{N}_\mathrm{N}}}
\newcommand{\NC}{\ensuremath{\mathcal{N}_\mathrm{C}}}
\newcommand{\NB}{\ensuremath{\mathcal{N}_\mathrm{B}}}
\newcommand{\FC}{\ensuremath{F_\mathrm{C}}}
\newcommand{\kk}{\ensuremath{k^\mathrm{(C)}}}
\newcommand{\kb}{\ensuremath{k^\mathrm{(B)}}}
\begin{document}
\title{Polarimetry and photometry of the peculiar main-belt object\\
7968 = 133P/Elst-Pizarro
\thanks{Based on observations made with ESO Telescopes at the Paranal Observatory under programme ID 079.C-0653 (PI=Tozzi)}}
       \author{
        S.~Bagnulo      \inst{1}
       \and
        G.P.~Tozzi      \inst{2}
       \and
        H. Boehnhardt   \inst{3}
       \and
        J.-B. Vincent   \inst{3}
       \and
        K.~Muinonen     \inst{4,5}
        }
\institute{
           Armagh Observatory,
           College Hill,
           Armagh BT61 9DG,
           Northern Ireland, U.K.\\
           \email{sba@arm.ac.uk}
           \and
           INAF - Osservatorio Astrofisico di Arcetri,
           Largo E. Fermi 5, I-50125 Firenze, Italy\\
           \email{tozzi@arcetri.astro.it}
           \and
           Max-Planck-Institut f\"{u}r Sonnensystemforschung,
           Max-Planck-Strasse 2,
           D-37191 Katlenburg-Lindau,
           Germany\\
           \email{boehnhardt@linmpi.mpg.de, vincent@linmpi.mpg.de}
           \and
           Observatory, PO Box 14,FI-00014 University of Helsinki, Finland\\
           \email{karri.muinonen@helsinki.fi}           
           \and
           Finnish Geodetic Institute, PO Box 15, FI-02431 Masala, Finland
          }
\authorrunning{S.\ Bagnulo et al.}
\titlerunning{Polarimetry and photometry of 133P/Elst-Pizarro}

\date{Received: 22 September 2009 / Accepted: 26 January 2010}
\abstract
{
Photometry and polarimetry have been extensively used as a
diagnostic tool for characterizing the activity of comets when they
approach the Sun, the surface structure of asteroids, Kuiper-Belt
objects, and, more rarely, cometary nuclei.}
{
133P/Elst-Pizarro is an object that has been described as either an
active asteroid or a cometary object in the main asteroid belt.
Here we present a photometric and polarimetric study of this object 
in an attempt to infer additional information about its origin.
}
{ 
With the FORS1 instrument of the ESO VLT, we have performed
during the 2007 apparition of 133P/Elst-Pizarro 
quasi-simultaneous photometry and polarimetry of its nucleus 
at nine epochs in the phase angle range $\sim
0\degr - 20\degr$. For each observing epoch, we also combined all
available frames to obtain a deep image of the object, to seek
signatures of weak cometary activity. Polarimetric data were
analysed by means of a novel physical interference modelling.
}
{ 
The object brightness was found to be highly variable over timescales $<
1$\,h, a result fully consistent with previous studies. Using the
albedo-polarization relationships for asteroids and our photometric
results, we found for our target an albedo of about 0.06--0.07 and a
mean radius of about 1.6\,km. Throughout the observing epochs, our deep
imaging of the comet detects a tail and an
anti-tail. Their temporal variations are consistent with an activity profile
starting around mid May 2007 of minimum duration of four
months. Our images show marginal evidence of a coma
around the nucleus. The overall light scattering behaviour
(photometry and polarimetry) resembles most closely that of F-type asteroids.
}
{}


\maketitle

\section{Introduction}
Light scattering of minor solar system bodies, such as asteroids, comets,
and Kuiper-Belt objects, plays a central role in determining
global physical parameters such as size and albedo, and the detailed
understanding of the surface structure such as micro structure and
single scattering albedo of the body surface \citep{Muinonen04}. This
applies to both remote sensing observations from Earth and to in situ
measurements during spacecraft encounters. Light scattering is studied
by means of photometric and linear polarimetric measurements obtained at
different phase angles, i.e., the angle between the Sun, the object,
and the observer.  The way in which photometry and linear polarization
change as a function of the phase angle helps us to characterize the
scattering medium, in both the case of solid surfaces and dust ejecta
around comets.

The phase angle range between 0\degr\ and about 30\degr is of
particular interest. At small phase angles ($\la 10\degr)$, the
linear polarization of small bodies in the solar system is usually
\textit{parallel} to the scattering plane, in contrast to what is
expected from the simple single Rayleigh-scattering or
Fresnel-reflection model, and the intrinsic brightness may increase
above the normal linear brightening with phase angle. Both phenomena
are a consequence of the surface micro structure and usually
attributed to the combined action of the shadowing effect and the
coherent backscattering of light. At larger phase angle values, the
value of the polarization minimum, the inversion angle at which the
polarization changes from being parallel to the scattering plane 
to becoming perpendicular to it, and the slope of the polarimetric curve
at the cross-over point differ for asteroids of different
surface taxonomy and may even be used to assess the albedo of
the object.

The systematic research and classification of the linear polarimetric and
photometric phase functions for the zoo of small bodies in the
solar system remains in its infancy, mostly because of a lack of good
coverage by measurements for the various objects types to be
considered (for instance, asteroids, cometary nuclei, Kuiper-Belt
objects, Trojans, minor satellites). Photometric and
  polarimetric techniques have been used to study asteroids
  \citep[e.g.,][]{Muinonen02}, Kuiper-Belt objects
  \citep[e.g.,][]{Boeetal04,Bagetal06,Bagetal08}, and the activity of comets
 while approaching the Sun,  \citep[e.g.][]{Penetal05}. In contrast, photometric
and polarimetric data of cometary \textit{nuclei} are
scarce in the literature.  Photometric phase functions of cometary
nuclei are available for fewer than a dozen of objects
\citep[see][]{Lametal04,Boeetal08,Lametal07a,Lametal07b,Lametal07c,
  Lietal07a,Lietal07b,Snoetal08,Tubetal08}. The polarization of
cometary nuclei is even less observed and studied than photometry. To
the best of our knowledge, only a single comet nucleus, that of 2P/Encke, has been observed in
polarized light \citep{Boeetal08}, over an extented phase-angle range.
Although the shape of the polarimetric curve of 2P/Encke resembles
to those of asteroids, the numerical values of
the polarization minimum, slope, and inversion phase-angle
differ remarkably from those of other small bodies (including
asteroids) in the solar system measured so far.

Here we present new observations of \centotrentatre, an object
classified as both a comet and an asteroid (7968 Elst-Pizarro) and seems
to be by its nature in between comets and asteroids. It is a small
km-size body \citep{Hsietal09} that belongs to the type of so-called main-belt comets \citep[MBC,][]{HsiJew06} of which only five
objects are known so far -- P/2005 U1 (Read), 176P/(118401) LINEAR,
  P/2008 R1 (Garradd), and P/2010 A1 (LINEAR) are the four
  others. It has an orbit rather typical of main-belt asteroids, for
which reason it is also designated as asteroid, is similar
to the Themis collision family, or a sub-family of
it, called the Beagle family, \citep{Nesetal08}.  It was discovered in 1996 by
\citet{Elsetal96} at the European Southern Observatory (ESO) in La
Silla (Chile); at that time it displayed a long thin dust tail, which
triggered its initial designation as a comet.  Although details
  of the activity's origin are as yet unknown, the activity producing
the tail seems to be recurrent and occur close to perihelion
passage, while the object may be inactive over the remainder of its orbit
around the Sun \citep{Hsietal04,Jewetal07}. Rather than a mostly
inactive comet, \centotrentatre\ is nowadays more commonly regarded as
an asteroid with an ice reservoir, which periodically sublimates,
with consequent material ejection \citep{Boeetal97, Hsietal04,
  Toth06}.

\section{Observations}\label{Sect_Observations}
Polarimetric and photometric observations of \centotrentatre\ were
obtained in service mode at the ESO Very Large Telescope (VLT) from
May to September 2007, with the FORS1 instrument \citep{Appetal98},
using broadband Bessell $R$ and $V$ filters. Linear polarization
measurements were obtained at nine different epochs, one before
perihelion (in May 2007), and the remaining eight after perihelion
(from July to September 2007). Each series of polarimetric
observations consisted of a 30\,s acquisition image obtained in the
$R$ Bessell filter without polarimetric optics, and a series of images
obtained with the half waveplate set at 12 to 24 position angles in
the range 0\degr -- 337.5\degr, in steps of 22.5\degr, both with $R$ and
$V$ Bessell filters. We also obtained photometric imaging in $R$ and
$V$ Bessell filters (i.e., without polarimetric optics) at the same
nine epochs (quasi-simultaneous to the polarimetric series).

We performed a preliminar inspection of the Line of Sight Sky
Absorption Monitor\footnote{Online data available at the ESO web site}
(LOSSAM) plots of each observing night showing the atmospheric
conditions on site, and we found that for all epochs, except one, sky
transparency was close to photometric at the time of the
observations. Night 29 to 30 August 2007, corresponding to our target
at phase-angle = 15.3\degr, was cloudy.  For polarimetry, the only
impact is in terms of a reduced signal-to-noise ratio, while
photometric measurements and tail length measurements obtained during
that night should not be considered reliable.  Some of the frames
obtained on 24 September could not be used because of background
objects overlapping the image of \centotrentatre.

\section{Data analysis}
Both polarimetric and photometric data, including acquisition frames
and science frames, were pre-processed in a similar way.  Frames were
bias-subtracted using a master bias obtained from a series of five
frames taken the morning after the observations, then divided by a
flat-field obtained by combining four sky flat images taken during
twilight with no use of polarimetric optics. From this point, we proceeded
with different strategies tailored to the specific analysis that we aimed
to perform.

\subsection{Deep imaging and tail analysis}\label{Sect_Deep_Imaging}
\begin{figure}
\scalebox{0.95}{
\includegraphics*[1.25cm,8.5cm][24cm,28.4cm]{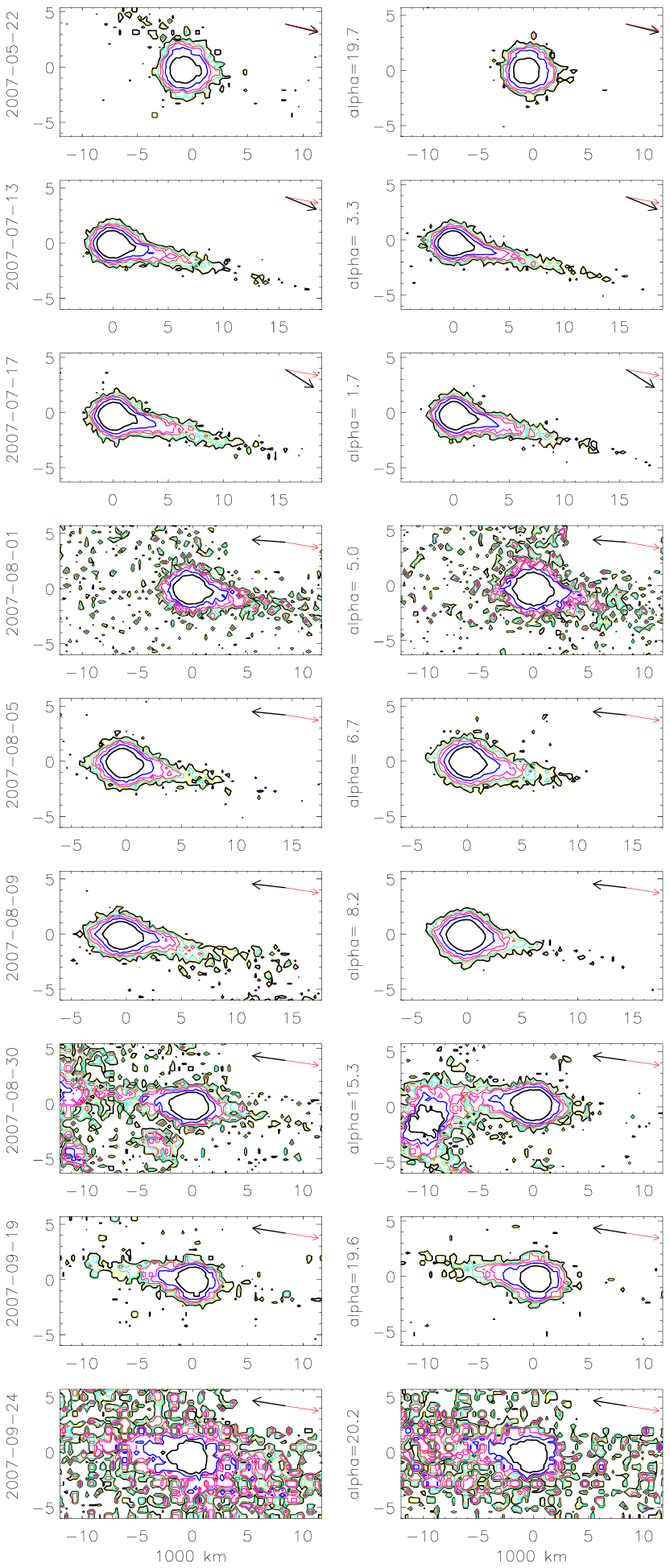}}
\caption{\label{Fig_Contours} Contour plots of $Af$ in $R$ and $V$
  Bessell filter (left and right columns,
  respectively). North is up, and east to the left. The black arrow
  and red arrow represent the direction of the Sun, and the negative
  target velocity as seen from the observer in the plane of the sky,
  respectively. The $x$ and $y$ axes are in units of $10^3$\,km. Contour levels
  are calibrated in $Af$ (for explanations see text). Contour
  levels correspond to $Af=7.0 \times 10^{-9}$, $8.3 \times 10^{-9}$, $1.0 \times 10^{-8}$,
  $1.25 \times 10^{-8}$, $1.65 \times 10^{-8}$, $2.5 \times 10^{-8}$, and
  $5 \times 10^{-8}$.}
\end{figure}
\begin{figure}
\scalebox{0.58}{
\includegraphics*[1.25cm,0.5cm][21cm,8.0cm]{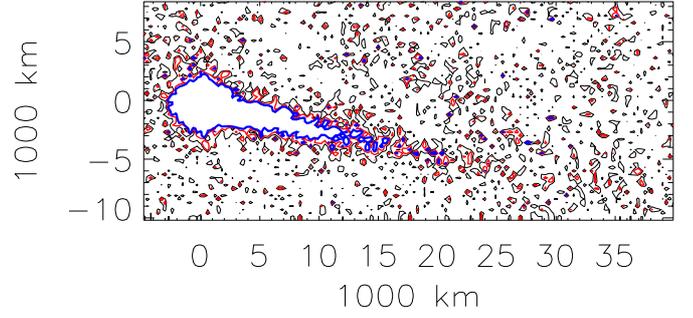}}
\caption{\label{Fig_July} Contour plots of $Af$ in $R$ Bessell filter for
the combined image obtained on July 17.
Contour levels correspond to
$Af = 2 \times 10^{-9}$, $4 \times 10^{-9}$, and $6 \times 10^{-9}$.
}
\end{figure}
Deep imaging of \centotrentatre\ was obtained by coadding the
polarimetric images as follows.  For each frame, we considered
separately the two images with opposite polarization that are split by
the Wollaston prism, from which we subtracted the background using
{\tt SEXTRACTOR}\footnote{http://astromatic.iap.fr/software/sextractor}
with the {\tt mesh} option. ``Full resolution'' background images
were checked to ensure that they were constant around the target,
and that background subtraction neither remove nor added small scale
features.
We then coadded the two images with opposite polarization for each of the
frames obtained at various positions of the retarder waveplate.
These frames were corrected for their corresponding airmass values.
Extinction coefficients, colour correction terms, and zeropoints were
obtained from the observations performed within the framework of the
FORS1 instrument calibration plan.  For the $R$ filter, we adopted the
values of $k_R=0.087 \pm 0.014$ and $k_{V-R} = -0.097 \pm 0.01$ for
the extinction and the colour correction, respectively; for the $V$
filter, we adopted $k_V=0.112 \pm 0.012$ and $k_{B-V} = -0.015 \pm
0.007$. For the colour indices of \centotrentatre\ we adopted
$V-R=0.41$ and $B-V=0.60$, which are close to solar colours.  However, the
instrument calibration plan does not include photometric calibration
of the images obtained with the polarimetric optics, and the
transmission function of the polarimetric optics is documented in
neither the FORS user manual nor the literature. We therefore
  performed a quick photometric calibration of the images obtained
  with the polarimetric optics by comparing the flux of the same
objects observed with and without polarimetric optics. We estimated
that the absorption of the polarimetric optics is about 0.60 and 0.54\,mag
in $R$, and $V$, respectively. Instrument zeropoints for the
polarimetric mode were thus calculated by subtracting 0.60 or 0.54
from the night zeropoints obtained within the context of the
instrument calibration plan in imaging mode. We note that this
  correction was obtained without accounting for a change in the
  extinction and colour coefficients caused by the polarimetric
  optics. The impact of this approximation should be minimal,
  relative to the uncertainties discussed in
  Sect.~\ref{Sect_Photometry}.

All frames obtained during a single epoch were average-combined using
a $\sigma$-clipping algorithm, adopting the pixel {\it median} value
as a center for clipping, and calculating the {\it average} of the
non-clipped pixels. These combined images
were then calibrated in $Af$, where $A$ is the albedo and $f$ the
filling factor, using Eq.~(1) of \citet{Tozetal07}
\begin{equation}
Af = 5.34 \times 10^{11} \left(r/d_{\rm p}\right)^2  
10 \times^{(m_\odot - \mathrm{ZP}_m)} \ \mathcal{N} \; ,
\end{equation}
where  $r$ is the heliocentric distance in AU, 
$m_\odot$ is the absolute magnitude of the Sun
in the considered filter, ZP$_m$ is the zeropoint in that filter,
and $d_{\rm p}$ is the CCD pixel scale in arcsec
(0.25\arcsec\ in our case), and $\mathcal{N}$ is the
number of electrons per pixel.
 
Images were finally magnified using a scale factor
\begin{equation}
s = 0.725\, \Delta \, d_{\rm p}\ ,
\end{equation}
where $\Delta$ is the geocentric distance in AU at the time of the
specific observation. In this way, the pixel coordinates where
converted into identical spatial coordinates expressed in 1000\,km.

Figure~\ref{Fig_Contours} shows the contour plots of the coadded
polarimetric images of \centotrentatre\ at the nine observing epochs,
i.e., from 38 days before to 87 days after perihelion.  

The image of 22 May 2007 shows a marginal indication of
a tail in the anti-solar direction. In all images obtained after
perihelion, we clearly detected one or two narrow
tails, either at position angles PA corresponding to the direction of
the Sun, at 180\degr\ from it, or in both directions. Since the
tails are seen in both $V$ and $R$ filters, we conclude that they
represent dust-reflected sunlight (and not so much -- or possibly not
at all -- from gas emission).  Table~\ref{Table_Tail} provides
geometric information about the two tail features measured from the
images.

\begin{table*}
\caption{\label{Table_Tail}
Geometric information about the dust tails of 133P/Elst-Pizarro.
}
\begin{center}
\begin{tabular}{crrcccc}
\hline
           &
\multicolumn{1}{c}{Phase}  &
\multicolumn{1}{c}{}       &
\multicolumn{2}{c}{Tail 1} &
\multicolumn{2}{c}{Tail 2} \\
Date       &
\multicolumn{1}{c}{Angle}  &
\multicolumn{1}{c}{PA$_\odot^{(1)}$}&
\multicolumn{1}{c}{PA$^{(2)}$}     &
\multicolumn{1}{c}{projected length$^{(3)}$} &
\multicolumn{1}{c}{PA$^{(2)}$}     &
\multicolumn{1}{c}{projected length$^{(3)}$}\\
yyyy mm dd &\multicolumn{1}{c}{(\degr)}&\multicolumn{1}{c}{(\degr)}  &\multicolumn{1}{c}{(\degr)}&\multicolumn{1}{c}{($10^{(3)}$\,km)}&\multicolumn{1}{c}{(\degr)}&\multicolumn{1}{c}{($10^{(3)}$\,km)}\\
\hline\hline 
2007 05 22 & 19.7 & 256.2 & 257 & -- & -- & --  \\
2007 07 13 &  3.4 & 248.1 & 255 & 18 & -- & --  \\
2007 07 17 &  1.7 & 237.7 & 256 & 17 & -- & --  \\
2007 08 01 &  5.0 &  85.7 & 257 & 12 & -- & --  \\
2007 08 05 &  6.7 &  84.2 & 260 & 12 & 81 &  3  \\
2007 08 09 &  8.2 &  83.4 & 258 & 12 & 82 &  4  \\
2007 08 30 & 15.3 &  81.9 & 262 & 7  & 82 &  6  \\
2007 09 19 & 19.6 &  81.0 & 261 & 4  & 81 & 10  \\
2007 09 24 & 20.3 &  80.7 & 261 & -- & 81 & 10  \\
\hline
\end{tabular}
\end{center}
(1) Position angle of the antisolar vector. \ \
(2) Position angle of the tail. \ \
(3) Approximate tail extension obtained using the contour level $Af = 7\,10^{-9}$ of 
Fig.~\ref{Fig_Contours}, averaged in the $R$ and $V$ filters.
\end{table*}

For comparison purpose, in Fig.~\ref{Fig_Contours} we adopted the same
contour levels for all epochs, $7\times10^{-9}$ being the one at the
smallest $Af$ value. However, we note that in all images, apart from one
obtained before perihelion, tails extend to a greater distance than
indicated by the $7\times10^{-9}$ contour level (although, below
this value, background noise becomes quite significant). The tail was
detected with the highest signal-to-noise ratio in the two images
obtained in July 2007, extending at least up to 25\,000\,km from the
photometric centre of the object, pointing toward the Sun. We note that
because of projection foreshortening, the measured length of the tail in
the sky compares to a much longer extension in space.  In the images
obtained on July 13, and in the regions at projected distance between
6\,000 and 20\,000\,km, we estimate that the scattering cross-section
of the tail (i.e., the projected surface multiplied by the albedo) is
about 0.27\,km$^2$ and 0.19\,km$^2$ in the $R$ and $V$ filter,
respectively.  In the images obtained on July 17 for the same
regions, we measure about 0.24\,km$^2$ and 0.21\,km$^2$ in the $R$ and
$V$ filter, respectively. Figure~\ref{Fig_July} shows the contour plot
for the combined image obtained on July 17 in the $R$ filter
setting for the lower contour level curve of value $Af = 2 \times 10^{-9}$.

In August, we detected, apart from the primary tail, a weak secondary tail,
which then prevailed in brightness over the first tail in the two
images obtained in September 2007. Hereafter, the primary one will be
referred to as ``Tail 1'', the secondary tail as ``Tail 2''.  Tail 1
points westward, i.e., close to opposition it is directed towards the
Sun. It appears as an anti-tail (a sun-ward pointing dust tail) in our
July 2007 image, and thereafter as a normal tail. It is brighter than
Tail 2 before about mid-August 2008, then fades away and is no longer
detectable in our last exposure series on 24 September 2007.  Tail 2 in
the eastern hemisphere always appears as an anti-tail. It is first
imaged in early August 2007 as a short eastward extension in the
isophote pattern in Fig.~\ref{Fig_Contours}, then brightens above
Tail 1 by the end of August, and remains detectable until the end of
our observations in late September 2007. The appearance of the two
tails in \centotrentatre\ resembles in terms of its behaviour the dust
phenomena observed in this object in 1995 and described by
\citet{Boeetal97}.

Using the Finson-Probstein (FP) code \citep{Beietal92,Bei1990}, we
could develop a qualitative and in part also quantitative
understanding of the dust activity of \centotrentatre. The FP code
allows one to calculate and display the so-called synchrone-syndyne
pattern of cometary dust tails, where synchrones represent the
location of the dust emitted from the nucleus at the same time and
syndynes represent the dust subject to the same solar radiation
pressure characterized by the $\beta_{\rm r}$ parameter, the ratio of
the force of solar radiation pressure to that of the gravity of the
Sun. During perihelion passage in 2007, the synchrone-syndyne pattern
of \centotrentatre\ shows a very dense and narrow grid of overlapping
lines, indicating a very narrow and spiky dust tail.  It also suggests
that the two observed tails are phenomena of projection caused by a
single dust tail, formed by the dust activity of the object that
lasted over a certain period around perihelion. Owing to the low
orbital inclination of \centotrentatre, the dust tail is seen almost
edge-on from Earth during the whole observing interval. Depending on
the observing dates, the FP calculations show that the two tails
represent different dust grain populations, emitted by the nucleus at
different time before the observing epoch and located on different
sides of the nucleus as projected onto the sky of the observer on
Earth. The projection-induced transition between Tails 1 and 2
happened shortly after opposition passage of the object on 20 July
2007.

From the images, we conclude that the dust activity of \centotrentatre\
was very low (or even absent) during our first observing
night (22 May 2007) 38 days before perihelion (top panels of
Fig.~\ref{Fig_Contours}). However, the presence of Tail 1 on 30 August
2007, and possibly on 19 Sept. 2007, suggests that significant dust
activity had started shortly (a few days) after 22 May 2007, since the
dust grains seen in this tail region must have been released by the
nucleus about 100 days or more before the date of observation, i.e., in
late May 2007. On the other side, the existence of Tail 2 in September
2007 is indicative of dust release by the nucleus that was still
ongoing by the time of our observations, i.e., almost 90 days after
perihelion passage.

The extension of Tail 2 from the nucleus allows us to estimate an
approximate maximum $\beta_{\rm r}$ value for grains in the most
distant region of this tail: we found a maximum $\beta_{\rm r}$ of
about 0.15, which is characteristic for instance of a few $\mu$m or ten $\mu$m
size grains from silicate or absorptive materials.  On 9 August 2007, the
width of Tail 1 is about 1000 km at a projected distance of 5000 km
from the nucleus indicating a slow (out-of-plane) expansion speed of
the dust of only about 1.5 m/s. The tail width on 13 and 17 July
  2007 at the same distance range (4000--5000\,km) was measured to be
  1750 and 2250\,km, respectively, which results in an expansion speed
  out of the orbital plane of about 1.45 ms$^{-1}$. Both results support
  the conclusions of a low expansion velocity of the dust grains in
  this object.

\begin{figure*}
\begin{center}
\scalebox{0.90}{
\includegraphics*[1.2cm,6.2cm][21cm,24.4cm]{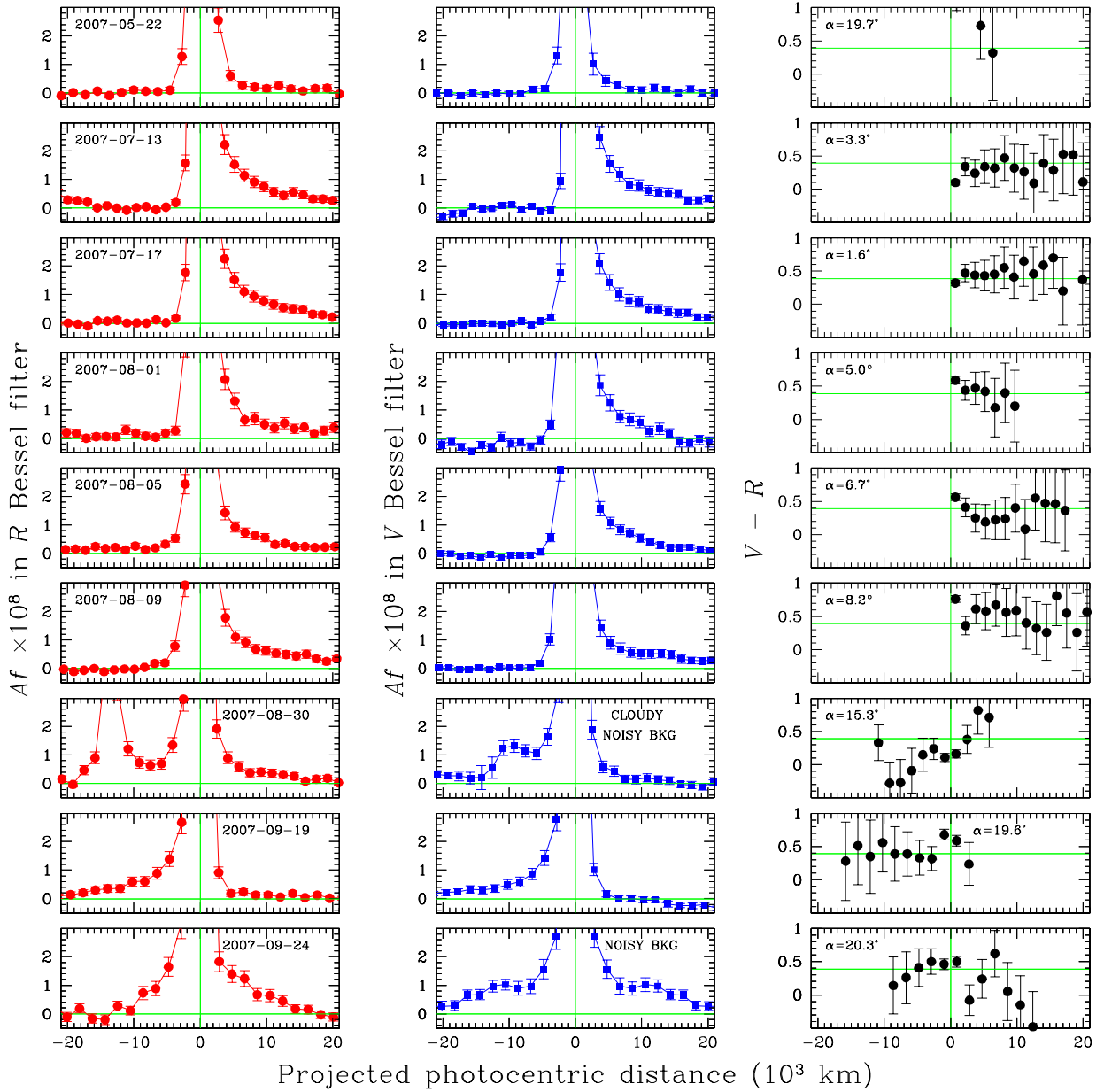}}
\end{center}
\caption{\label{Fig_Tail_Af} Tail $Af$ measured along the direction
  identified by the tail, with respect to the photometric center, and
  tail colour indices $V-R$. In the left and middle column panels,
  green lines show the zero axes (corresponding to the nucleus
  position). In the right panels, they correspond to the colour index
  derived for the nucleus.}
\end{figure*}

Finally, to obtain tighter constraints of the tail(s) brightness, we
calculated the average $Af$ at different distances along the direction
identified by the tail(s). We measured the average flux in rectangular
areas of 20 pixels, five along the direction identified by the tail,
and four in the direction perpendicular to it.\footnote{We note that the
measured tail $Af$ values may be possibly overestimated by up to $10$\,\%
because of a faint coma (see Sect.~\ref{Sect_Search_Coma}).}
Figure~\ref{Fig_Tail_Af} shows the $Af$ in the $R$ and $V$ filters,
and the colour index $V-R$, versus distance from the photometric
centre of the object. The points at positive distances refer to Tail
1, and those at negative distances refer to Tail 2. The $V-R$ points
of the rightmost panels were plotted only when flux was detected in
both filters at the minimum 2\,$\sigma$ level.  The $V-R$ colour of the
tails is about 0.4. It is mostly constant along the tail axes, and
close to a neutral intrinsic colour typical of a flat spectrum in the
visible. It is also in good agreement with the $V-R$ colour of the
nucleus itself (see Sect.~\ref{Sect_Photometry}). The neutral colour
of the dust tails is compatible with light scattering by grains that
were much larger than the wavelengths of the filter measurements. It
also implies that the surface material and the dust released by
\centotrentatre\ is not ``red'' as frequently observed for cometary dust
and considered indicative of the space weathering effects on the surface
materials.

\subsection{Searching for coma activity}\label{Sect_Search_Coma}
\begin{figure}
\scalebox{0.65}{
\includegraphics*[1cm,6.5cm][26cm,25cm]{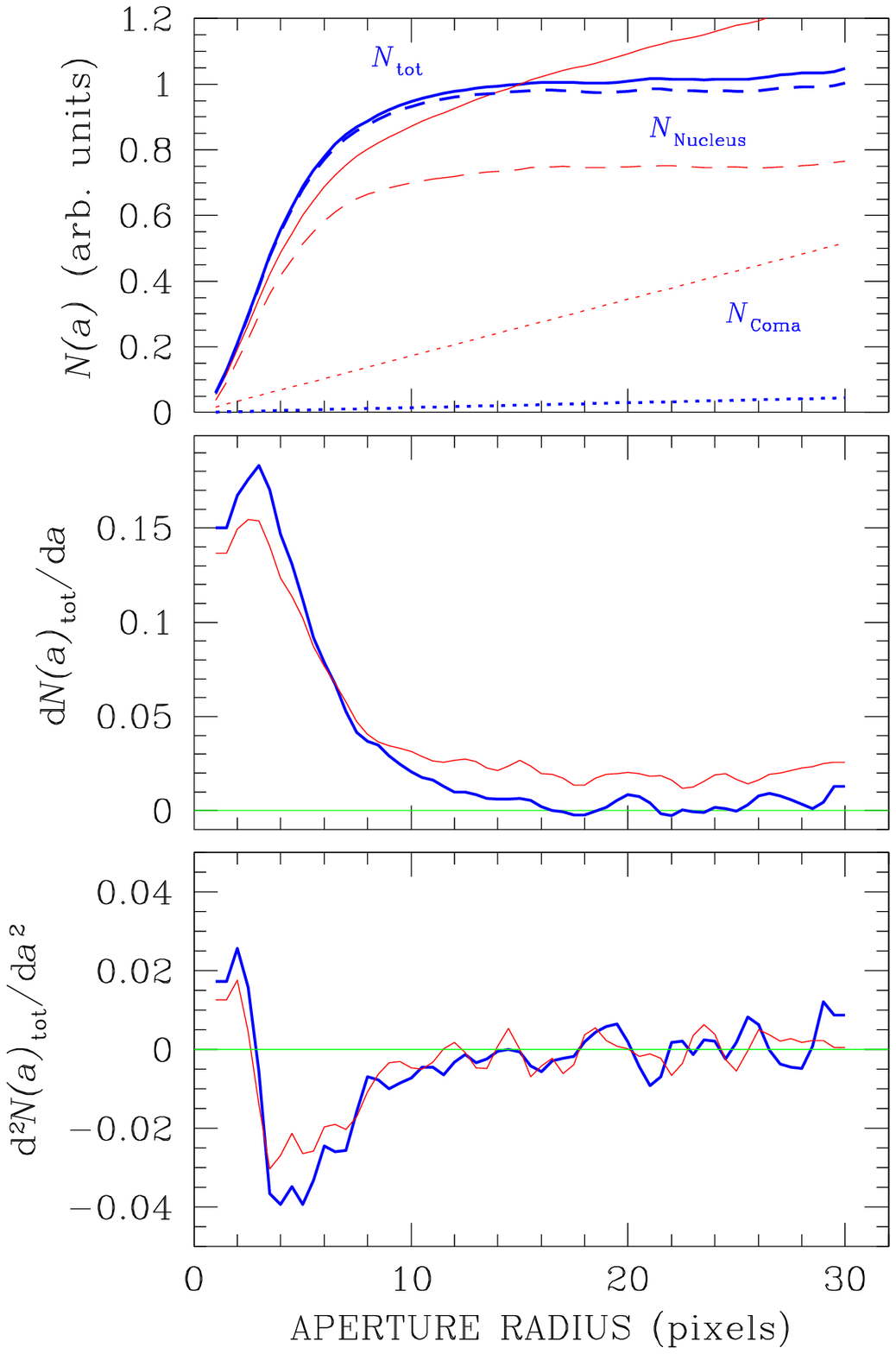}}
\caption{\label{Fig_Ref_Star}
From top to bottom: $\mathcal{N}$, $\mathcal{N}'$, and $\mathcal{N}''$
profiles for a background reference star (blue thick lines) and for the 
object \protect\centotrentatre\ (red thin lines), observed on 22 May 2007
in the $R$ filter.
All profiles are plotted versus the aperture measured in pixels (for
\centotrentatre, on 22 May 2007, a 0.25\arcsec\ pixel corresponds to 365\,km).
Together with the $\Ntot(a)$ profile (solid lines), the top
panel shows also the two contributions $\NN(a)$ (dashed lines),
and $\NC(a)$ (lower dotted lines). 
}
\end{figure}
\begin{figure}
\scalebox{0.45}{
\includegraphics*[1cm,5.7cm][21cm,25cm]{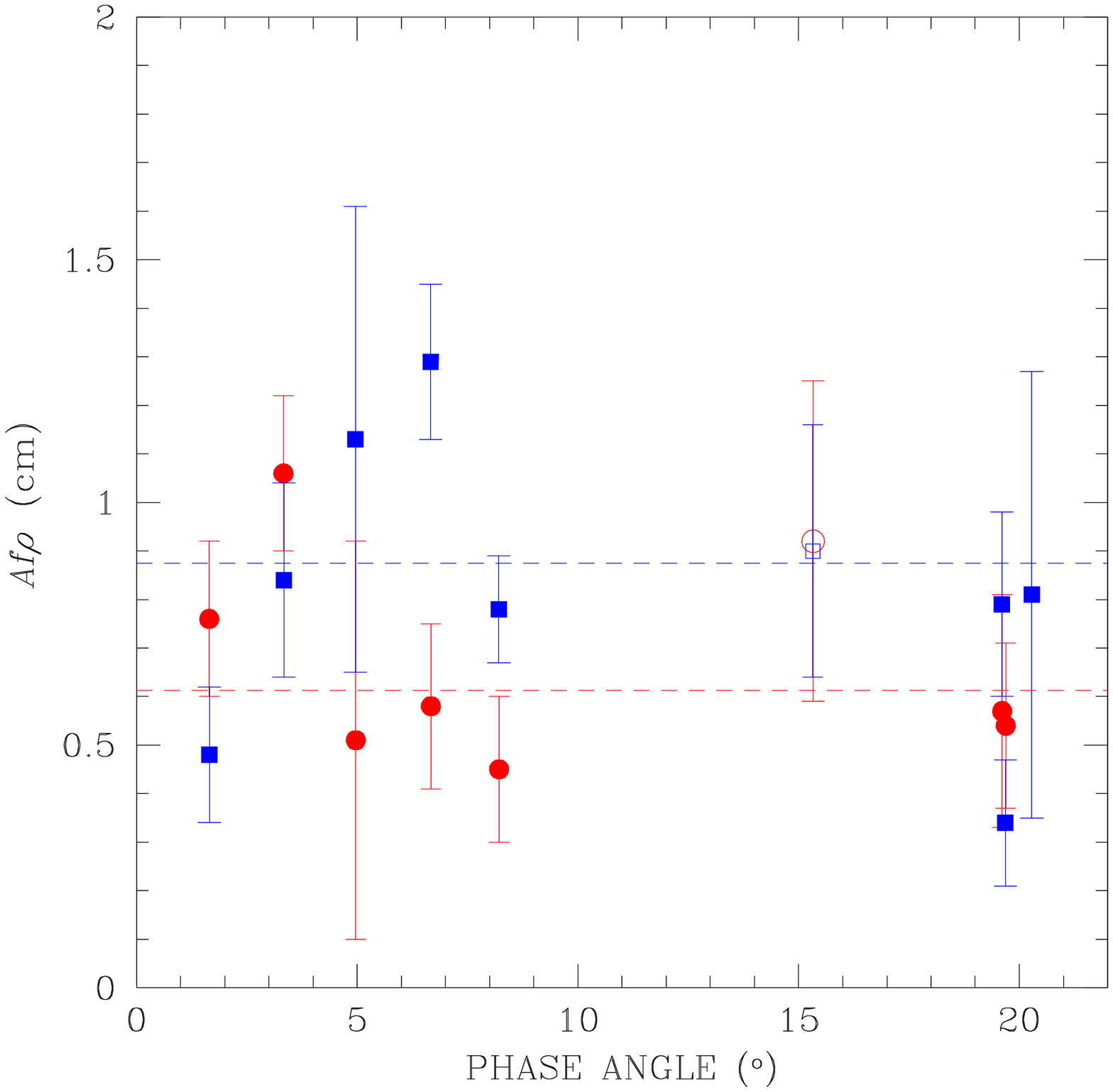}}
\caption{\label{Fig_Coma} $Af\rho$ for the coma on 22 May 2007. Red circles refer
to $R$ filter, blue squares to $V$ filter. Dashed lines mark the
average $Af\rho$ values in the two filters, obtained neglecting the points
marked with empty symbols.
}
\end{figure}
We provide a formal description of our approach
to evaluate coma contribution from our measurements. This is based on
a strategy originally developed by \citet{TozLic02} and
\citet{Tozetal04}. Thereafter, we describe the results of the
analysis performed on the combined polarimetric images of
\centotrentatre.

We assume that the number of detected electrons e$^-$ of the object
per unit of time within a circular aperture of radius $a$, $\Ntot(a)$, is the
sum of the contribution of the nucleus, \NN, plus the contribution of
the coma, \NC\, and, possibly, a spurious contribution \NB, due to
non-perfect background subtraction
\begin{equation}
\Ntot(a) = \NN(a) + \NC(a) + \NB(a) \ ,
\label{Eq_e-photometry}
\end{equation}
where $a$ is the radius of the aperture in pixels. Following
\citet{AHeetal84}, the flux of a (weak) coma around the nucleus in
a certain wavelength band can be written as
\begin{equation}
\FC = A f  \left(\frac{\rho}{2 \Delta}\right)^2 \frac{1}{r^2} {F_\odot}\; ,
\label{Eq_Af}
\end{equation}
where $A$ is the mean albedo (unitless), $f$ is the filling factor
(unitless), $\Delta$ is the geocentric distance and $\rho$ is the
projected distance from the nucleus (corresponding to the aperture in
which the flux \FC\ was measured), expressed in identical units, $r$ is
the heliocentric distance expressed in AU, $F_\odot$ is the solar flux
at 1\,AU integrated in the same band as \FC, and convolved with the
filter transmission curve. Equation~(\ref{Eq_Af}) applies to the
reflected light of the dust coma, which in the visible clearly
dominates over contributions from gas emission. Based on the
hypothesis of uniform and isotropic ejection of dust at constant
velocity, \citet{AHeetal84} note that the product $A f \rho$ is
constant with $\rho$. The background is assumed to be constant in the
small region measured around the object, hence, linearly proportional
to the aperture area.  Equation~(\ref{Eq_e-photometry})
can thus be written
\begin{equation}
\Ntot(a) = \NN(a) + \kk a + \kb a^2 \ ,
\label{Eq_smart-photometry}
\end{equation}
where \kk\ and \kb\ are terms independent of $a$.
We then trivially obtain
\begin{equation}
\begin{array}{rcrcl}
\Ntot'(a) &=& \frac{\mathrm{d}}{\mathrm{d} a}\Ntot(a)     &=& \NN'(a) + \kk + 2a \kb \ , \label{Eq_first}\\[2mm]
\Ntot''(a) &=& \frac{\mathrm{d}^2}{\mathrm{d} a^2}\Ntot(a) &=& \NN''(a) + 2 \kb       \label{Eq_second} \; .\\
\end{array}
\end{equation}
Plotting $\Ntot''(a)$ versus $a$ allows one to estimate whether
background subtraction is optimum, as at large apertures compared to
the comet image -- barring the presence of background objects --
$\Ntot''(a)$ should converge toward the $\kb$ value,
which for optimum sky subtraction should be zero.

Assuming that the nucleus is a point source, the term $\NN'(a)$ is the
point spread function, and, for apertures sizes that are large
compared to the seeing, should tend to zero. If an extended coma
is present, we expect a contribution from the term  $\NC'(a)$ constant
with $a$.  The coma contribution can then be evaluated by measuring
the constant \kk\ as a weighted average of $\Ntot'(a)$ in the aperture
interval $[a_{i_1},a_{i_2}]$, with $a_{i_1} < a_{i_2}$.
The \kk\ value can finally be used in Eq.~(\ref{Eq_smart-photometry})
to distinguish, for each aperture value, the flux due to the coma and that
due to the nucleus. In particular, the former one
increases linearly with aperture, and the latter should appear constant
at the aperture values at which $\Ntot'(a)$ is constant, and which
were used to determine \kk.

The nucleus contribution \NN\ can then be transformed to magnitude
$m$, following the standard recipes for photometry.  The coma
contribution \NC\ can be transformed into astrophysically meaningful
terms using the quantity $A f \rho$ introduced by
\citet{AHeetal84}, using the formula
\begin{equation}
A f \rho = 1.234 \times 10^{19} \ 10^{0.4\,(m_\odot - \mathrm{ZP}_m)} \
           r^2 \ \left(\frac{\Delta}{d_{\rm p}}\right) \ \kk\ .
\label{Eq_Afrho}
\end{equation}
where $r$ and $\Delta$ are measured in AU, 
$d_{\rm p}$ in arcsec per pixel, $\kk$ in e$^-$ per pixel, and
$A f \rho$ is obtained in cm.

To test this algorithm, we used the frames obtained on 22 May 2007,
when \centotrentatre\ exhibited the least evidence of tail
activity, and its image appeared relatively isolated from background
objects. We applied our algorithm to both \centotrentatre\ and an
isolated background star of similar brightness. Suitable comparison
objects are extremely rare in our images. During a series
of exposures, differential tracking causes background objects to
shift away from the field of view limited by the 22\arcsec\
wide strip mask used in polarimetric mode. However, since
\textit{individual} exposures were short (between 30\,s and 100\,s),
star trailing was limited always to less than 1 pixel
size, compared to a typical seeing of 4 pixels.

The results of our test are illustrated in Fig.~\ref{Fig_Ref_Star},
which shows that for the background star $\NC'(a)$ tends to zero for
$a \ga 20$\,pixels, while for \centotrentatre\ $\NC'(a)$ converges
to a positive constant value. In Fig.~\ref{Fig_Ref_Star}, the
constant used to disentangle the $\NN(a)$ and $\NC(a)$ profiles was
calculated by interpolating with a constant term the $\Ntot'(a)$
profiles (shown in the middle panel) between 22 and 28 pixels. For
both the background star and \centotrentatre, all profiles were
normalised imposing that the fluxes integrated within a 15 pixel
(=3.75\arcsec) aperture are equal to 1.

The most critical issue is that the coma region (if a coma is present
at all) is contaminated by the tail contribution, which so far we have
implicitly neglected. Therefore, we transformed our images
into polar coordinates ($a,\theta$) and removed the regions
within those azimuth ranges that were clearly contaminated by tail(s)
or background sources. The flux pertaining to a certain annulus
$[a,a+da]$ was then calculated by integrating the pixel values at the
various $\theta$ values contaminated by netiher tail nor background
sources, multiplied by a factor $\ge 1$ to account for the image
trimming. The errorbars in the $\NN'$ profiles were estimated by
associating with the measured flux the standard deviation of the
distribution of the fluxes at the various $\theta$ values.  Finally,
we applied an algorithm similar to that described in the case of
cartesian coordinates. The coma
contribution was measured in circular regions between 4500 and
9000\,km from the nucleus photo-centre.

The results of this analysis, shown in Fig.~\ref{Fig_Coma}, are
consistent with there being a coma with $A f \rho$ $\la 1$\,cm
detected at a 2-5\,$\sigma$ level. This result could well be ascribed
to stray light in the instrument (which may mimic a diffuse halo).  We therefore 
repeated the same analysis
on the images of a number of background stars at various observing
epochs. We found that the ratios $\NC(a)/\Ntot(a)$ were generally
substantially higher for \centotrentatre\ than for background objects.
In conclusion, our data exclude the presence of a coma with $Af\rho
\ga 3$\,cm, but certainly do not allow us to rule out the possibility
that a faint coma, with $Af\rho$ of the order of 1\,cm or smaller,
exists. Data shown in Fig.~\ref{Fig_Coma} marginally suggest a change 
of the colour index, but reaching a firm conclusion in that respect
would require higher signal-to-noise ratio data.

This value compares to a dust production rate of the order of 100
g\,s$^{-1}$ \citep{Boeetal08}, which would be one of the lowest level
ever measured for a comet, although lasting most likely for several
months around perihelion. Despite its large uncertainty (order of
  one magnitude), this dust production rate is still higher than that
  obtained by \citet{Hsietal04}, which could be caused by temporal
  variability in the nucleus activity and/or measurement and modeling
  errors (in both datasets).  The most logical explanation of the almost
absent coma and the narrow dust tails is low
nucleus activity and the small terminal expansion velocity of the dust
grains after release by the nucleus of \centotrentatre. Small
dust expansion velocities can be concluded from estimations of
the tails. Hence, the dust coma is weak and confined very much to the
near-nucleus region, which makes it difficult to detect in the
atmospheric seeing disk of the latter.

\subsection{Nucleus photometry}\label{Sect_Photometry}
\begin{figure}
\scalebox{0.45}{
\includegraphics*[0.7cm,5.8cm][24cm,25.2cm]{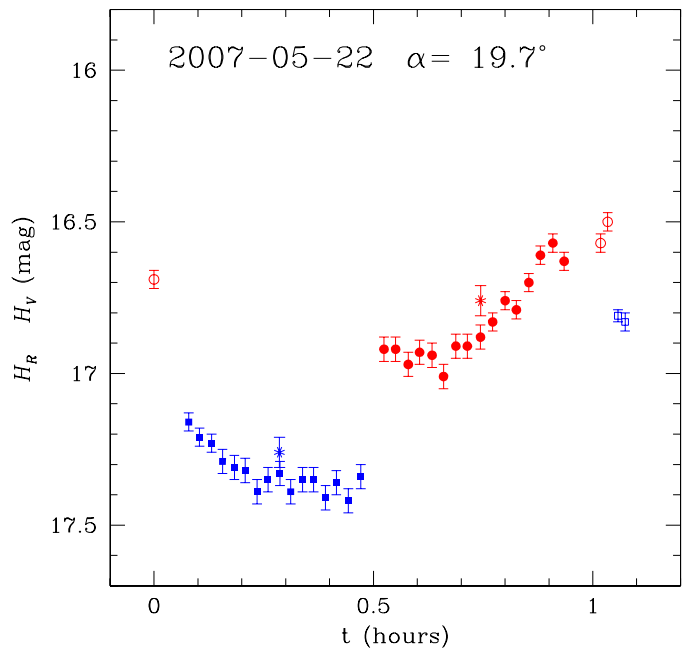}}
\caption{\label{Fig_Light_Curve}
  Photometry in the $R$ (red circles) and $V$ (blue squares) filters
  for \protect\centotrentatre\ as function of time, obtained on
  22 May 2007. Filled symbols refer to the photometry obtained from the
  polarimetric images, empty symbols refer to the photometry
  obtained from the images with no polarimetric optics in. Asterisks
  show the photometry obtained from the averaged polarimetric images. 
  Point at $t=0$ is obtained from the acquisition image of the
  polarimetric series.
}
\end{figure}

For each observing night, our data set typically consists of an
acquisition image in $R$, a series of 12 to 32 polarimetric images in
$V$, followed by a series of 12 to 32 polarimetric images in $R$, and
a series of two photometric images in $R$ and two in $V$. Images were
taken approximately two or three minutes apart, except for the
photometric series, which was occasionally taken ten to thirty minutes
before or after the polarimetric series. For each epoch, the set of
photometric data consists of the acquisition
image of the polarimetric series in $R$, the averaged frames obtained
from all polarimetric images in $R$ and $V$, and images obtained in
$R$ and $V$ with no polarimetric optics. 
\begin{table*}
\caption{\label{Tab_Photometry}
Photometry of comet 133P/Elst-Pizarro.
}
\begin{center}
\begin{tabular}{crrcccccc}
\hline
Date                            &
\multicolumn{1}{c}{Phase}       & 
\multicolumn{1}{c}{$r^{(1)}$}       &
\multicolumn{1}{c}{$\Delta^{(2)}$}  &
\multicolumn{3}{|c}{$H_R$ ($\alpha$) (mag)} &
\multicolumn{2}{|c}{$H_V$ ($\alpha$) (mag)} \\
(yyyy mm dd)                 &
\multicolumn{1}{c}{(\degr)}  &
\multicolumn{1}{c}{(AU)}     &
\multicolumn{1}{c}{(AU)}     &
acq.$^{(3)}$          &
aver. pol.$^{(4)}$    &
photom.$^{(5)}$       &
aver. pol.$^{(4)}$    &
photom.$^{(5)}$      \\
\hline\hline
2007 05 22 & 19.69  & 2.6467 & 2.0129 & 16.69 &  16.76  &16.53 & 17.26 & 16.82  \\ 
2007 07 13 &  3.33  & 2.6424 & 1.6329 &       &  15.76  &15.62 & 16.11 & 15.94  \\ 
2007 07 17 &  1.66  & 2.6429 & 1.6282 & 15.57 &  15.75  &15.53 & 16.04 & 15.88  \\ 
2007 08 01 &  4.95  & 2.6455 & 1.6466 & 15.68 &  15.86  &15.65 & 16.22 & 16.13  \\ 
2007 08 05 &  6.68  & 2.6465 & 1.6619 & 15.91 &  15.69  &15.92 & 16.26 & 16.40  \\ 
2007 08 09 &  8.20  & 2.6475 & 1.2794 & 15.95 &  15.82  &15.91 & 16.27 & 16.35  \\ 
2007 08 30 & 15.32  & 2.6548 & 1.8342 &(16.19)& (16.34) &(16.51)&(16.33)&(16.82)\\ 
2007 09 19 & 19.61  & 2.6646 & 2.0521 & 16.46 &  16.25  &16.36 & 17.00 & 16.94  \\ 
2007 09 24 & 20.29  & 2.6674 & 2.1130 & 16.42 &         &16.74 & 16.63 & 17.22  \\ 
\hline
           & 0.00   &        &        &\multicolumn{3}{c}{$15.50 \pm 0.05$}&\multicolumn{2}{c}{$15.88 \pm 0.07$} \\
\hline
\end{tabular}
\end{center}

\noindent
(1) Helio-centric distance to the object. \ \
(2) Geo-centric distance to the object. \ \
(3) Magnitude measured in the acquisition images. \ \
(4) Magnitude measured in the averaged polarimetric frames. \ \
(5) Magnitude measured in the frames obtained with no polarimetric optics in. \ \ 
The last row refers to the brighntess at phase-angle 0\degr\ 
extrapolated with a linear best-fit. \ \ \ Blanks refer to frames where the image of 
\centotrentatre\ was contaminated by background objects. Values given in
parenthesis should be considered as lower limits because obtained during a
cloudy night.
\end{table*}


For all images, we used aperture photometry, adopting an aperture
radius of 12 pixels ($=3\arcsec$). Our nucleus brightness measurements
are contaminated by the emission of a tail and (possibly) a coma.
Tail emission was roughly taken into account by subtracting, from the
flux integrated in a circular aperture of radius $a$ (in pixels), the
amount corresponding to $Af=2.5 \times 10^{-8}$ per pixel, multiplied by an
area of $4\,(a-2)$ pixels, both for the $R$ and the $V$ filter, for
all observing epochs apart from the one before perihelion, when the tail was
very faint. For the adopted 12\,pixel aperture, this corresponds
typically to a 0.1\,mag correction. Coma contribution was subtracted
using the results of the previous section.  For a 12\,pixel aperture,
this corresponds to a correction generally of the order of $0.1$ mag.
Finally, the apparent magnitudes were converted into normalised
magnitude using 
\begin{equation}
H_m(\alpha) = m - 5\,\log (r\,\Delta) \; ,
\label{Eq_Reduced_Mag}
\end{equation}
where $r$ and $\Delta$ (the heliocentric and the geocentric
distances), are expressed in AU, and $\alpha$ is the phase-angle.

For each observing epoch, we finally obtained a light curve with a
typical time baseline of the order of 1\,h 15\,m to 1\,h 30\,m.  At
all the observing epochs, we could clearly note that the object
photometry varyed within a short timescale ($\le 1$\,h). The
most extreme case, shown in Fig.~\ref{Fig_Light_Curve}, is night
22 May 2007, during which we measured an amplitude of about 0.5\,mag
and 0.6\,mag in $R$ and $V$, respectively, to be compared to a
variability of $\la 0.02$\,mag measured for a nearby background
star. In the observations obtained on 19 September 2007, which have the
shortest time baseline of our dataset (1.1\,h), \centotrentatre\
exhibited the smallest brightness variability $\la 0.2$\,mag in both
filters.  While in some cases, the variability of \centotrentatre\ may
be explained by the close presence of background objects or by changes
in the sky transparency, in most of the cases the observed variability
is intrinsic to the object.  \citet{Hsietal04} estimated a light curve
amplitude of the order of 0.4 magnitude in $R$ filter, with a 3.471\,h
period. Earlier observations reported substantially larger variability
\citep[see][and references therein]{Hsietal04}, which may indicate
changes of the nucleus aspect angle along the orbit (although
  some results reported previously may be
  affected by poor calibration).

We note that our observational dataset may possibly be suited to a
detailed period analysis based on differential photometry, which is
beyond the scope of this paper. The aim of our photometric
measurements is to determine, by extrapolating of the
brightness measured at various phase angles, an estimate of the
absolute brightness of the object nucleus at phase angle
0\degr. With this in mind, the uncertainty introduced by using
simple aperture photometry (as opposed to the PSF fitting method), and by
a less than optimal photometric characterization of the observing
nights, is negligible compared to the variability of the
object due to its rotation.  The major contribution to the errorbars
comes from the uncertainty in the zeropoints, and in both the tail and coma
contributions, the sum of which was finally estimated to be 0.1\,mag
for all points.


\begin{figure}
\scalebox{0.45}{
\includegraphics*[0.75cm,5.5cm][21cm,25cm]{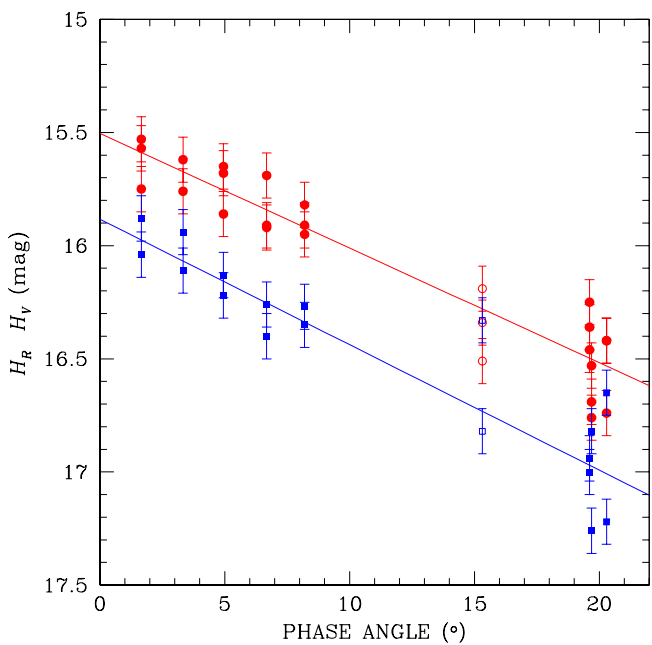}}
\caption{\label{Fig_Photometry}
Photometry in the $R$ (red circles) and $V$ (blue squares)
filters for \protect\centotrentatre\
as function of the phase angle. Solid lines show the corresponding 
linear best-fits. Points that were not used to calculate the best-fit 
are represented with empty symbols.
}
\end{figure}

Our results are given in Table~\ref{Tab_Photometry} and plotted in
Fig.~\ref{Fig_Photometry}.  The slopes of the brightness curves are
$0.052 \pm 0.004$\,mag/deg in $R$, and $0.055 \pm 0.005$\,mag/deg in
$V$, indicating a phase darkening coefficient in the range typical of
cometary nuclei. The extrapolated average brightnesses at phase angle
= 0\degr\ are $15.50 \pm 0.05$ and $15.88 \pm 0.07$ in the $R$ and $V$
filters, respectively, resulting in an average colour index
$V-R=0.38$, which is equivalent to an average spectral gradient of
+2\%/100\,nm in the $V$ and $R$ wavelength range, i.e., the intrinsic
color of \centotrentatre\ appears to be neutral, and a flat solar
reflectance spectrum in the visible is expected, at least beyond about
500\,nm.

The absolute brightness of \centotrentatre\ in $R$ and $V$ filter
clearly increases towards zero phase angle, but no indication of an
opposition brightening is found. Deviations from linearity may be caused by
measurement errors and (mostly) rotation variations of the
elongated body \citep[][give a minimum axis ratio of 1.45 and a
rotation period of 3.471\,h]{Hsietal04}.  The absolute brightnesses at
zero phase-angle correspond to an average equivalent radius of 1.6\,km
(range is 1.64--1.72\,km in $R$ to 1.47--1.57\,km in $V$), assuming the
0.07 and 0.06 albedo values (in $R$ and $V$, respectively) obtained
from our polarimetric measurements.

We finally compare our results with those of previous studies, noting
that previous photometric measurements were obtained more than four
months \citep{Boeetal97} and more than seven months \citep{Hsietal04}
after perihelion in 1996 and 2002, respectively, while the
measurements presented in this work cover the orbit arc from about one
month before to three months after perihelion in 2007.

\citet{Hsietal04} and \citet{Jewetal07} found a phase darkening
coefficient $\beta_{\rm d} = 0.044 \pm 0.007$\,mag/deg and an average
$V-R$ color of $0.42 \pm 0.03$, which are fully consistent with our own
measurements. We note that colour indices obtained from
quasi-simultaneous images in $R$ and $V$ vary between 0.29 and
0.48, which is marginally larger than reported by
\citet{Hsietal04}, who measured $V-R$ varying between 0.35 and
0.49. 

The nucleus size has been previously estimated (neglecting possible
contamination from coma and tails) to be about $2.5 \pm 0.1$\,km
assuming an albedo of 0.04, which is also consistent with our values
when applying identical albedo parameters. 
\citet{Hsietal09} give an equivalent radius of the nucleus of
\centotrentatre\ of $1.9\pm0.3$\,km for a geometric albedo of 0.05.

We note that the variation amplitudes of 0.5 to 0.6\,mag seen in our
photometric data over time intervals between 1\,h and 1.5\,h, compare to a
minimum aspect ratio (ratio of the long to short axes lengths for a
prolate ellipsoid rotating about its small axis) of about 1.6 to
1.7. These values are slightly larger than that measured by
\citep{Hsietal04}, which could be due to the different aspect angles of the
varying nucleus cross-section along the orbit of the object.

\subsection{Imaging polarimetry}\label{Sect_Polarimetry}
\begin{table*}
\caption{\label{Tab_ElPiza}
Polarization measurements of \centotrentatre.
}
\begin{center}
\begin{small}
\begin{tabular}{ccrrrrcrr@{\,$\pm$}lr@{\,$\pm$}l}
\hline\hline
\multicolumn{1}{c}{Date}   &
\multicolumn{1}{c}{UT$^{(1)}$}  &
\multicolumn{1}{c}{Exp}    &
\multicolumn{1}{c}{PA$^{(2)}$}     &
\multicolumn{1}{c}{Phase Angle}  &
\multicolumn{1}{c}{$t-t_0^{(3)}$}   &
\multicolumn{1}{c}{Filter} &
\multicolumn{1}{c}{Ap.$^{(4)}$}&
\multicolumn{2}{c}{\pq}    &
\multicolumn{2}{c}{\pu}    \\
\multicolumn{1}{c}{yyyy mm dd}   &
\multicolumn{1}{c}{hh:mm}  &
\multicolumn{1}{c}{(s)}    &
\multicolumn{1}{c}{($^\circ$)}     &
\multicolumn{1}{c}{($^\circ$)}     &
\multicolumn{1}{c}{(days)} &
\multicolumn{1}{c}{}       &
\multicolumn{1}{c}{pix.}   &
\multicolumn{2}{c}{(\%)}   &
\multicolumn{2}{c}{(\%)}  \\
\hline

 2007 05 22 &09:44 &  960& 256.160& 19.690 & $-$37.66 &$R$& 7.0&$  0.96$&$ 0.59$&$  0.10$&$ 0.59$\\
 2007 05 22 &09:17 &  960& 256.162& 19.693 & $-$37.68 &$V$& 7.0&$  0.85$&$ 0.55$&$ -0.01$&$ 0.56$\\[2mm]
 2007 07 13 &07:43 &  960& 247.877&  3.331 & $+$14.25 &$R$& 5.0&$ -0.75$&$ 0.16$&$ -0.39$&$ 0.16$\\
 2007 07 13 &07:13 &  960& 247.905&  3.340 & $+$14.23 &$V$& 5.0&$ -1.27$&$ 0.15$&$ -0.10$&$ 0.16$\\[2mm]
 2007 07 17 &07:05 &  960& 237.251&  1.646 & $+$18.22 &$R$& 7.0&$ -0.53$&$ 0.16$&$ -0.16$&$ 0.16$\\
 2007 07 17 &06:37 &  960& 237.360&  1.654 & $+$18.21 &$V$& 5.5&$ -0.53$&$ 0.15$&$ -0.04$&$ 0.15$\\[2mm]
 2007 08 01 &02:47 &  960&  85.710&  4.969 & $+$33.05 &$R$& 5.5&$ -1.17$&$ 0.31$&$  0.02$&$ 0.29$\\
 2007 08 01 &02:19 &  960&  85.720&  4.960 & $+$33.03 &$V$& 7.0&$ -0.50$&$ 0.39$&$  0.09$&$ 0.39$\\[2mm]
 2007 08 05 &04:58 & 1280&  84.169&  6.671 & $+$37.14 &$R$& 5.0&$ -1.28$&$ 0.14$&$  0.26$&$ 0.15$\\
 2007 08 05 &04:27 & 1280&  84.174&  6.662 & $+$37.12 &$V$& 6.0&$ -1.48$&$ 0.15$&$ -0.14$&$ 0.15$\\[2mm]
 2007 08 09 &01:09 &  960&  83.378&  8.215 & $+$40.98 &$R$& 5.0&$ -1.22$&$ 0.17$&$  0.12$&$ 0.17$\\
 2007 08 09 &00:42 &  960&  83.381&  8.207 & $+$40.96 &$V$& 5.0&$ -1.53$&$ 0.18$&$  0.02$&$ 0.18$\\[2mm]
 2007 08 30 &00:43 & 1280&  81.907& 15.330 & $+$61.96 &$R$& 6.0&$ -0.28$&$ 0.46$&$  0.03$&$ 0.48$\\
 2007 08 30 &00:13 & 1280&  81.908& 15.324 & $+$61.94 &$V$& 6.0&$ -0.51$&$ 0.29$&$ -0.02$&$ 0.30$\\[2mm]
 2007 09 19 &02:16 & 1200&  80.962& 19.611 & $+$82.02 &$R$& 6.0&$  0.33$&$ 0.37$&$ -0.32$&$ 0.36$\\
 2007 09 19 &01:48 & 1200&  80.963& 19.608 & $+$82.00 &$V$& 5.0&$  0.99$&$ 0.39$&$  0.08$&$ 0.37$\\[2mm]
 2007 09 24 &01:01 & 1280&  80.655& 20.288 & $+$86.97 &$R$& 7.0&$  1.63$&$ 1.13$&$  0.27$&$ 1.17$\\
 2007 09 24 &00:28 & 1280&  80.656& 20.285 & $+$86.95 &$V$& 6.5&$  1.94$&$ 0.90$&$ -0.64$&$ 0.88$\\
\hline
\end{tabular}
\end{small}
\end{center}
(1) Date and UT refers to the middle of the exposure series. \ \
(2) Position angle of the scattering plane. \ \
(3) Time from perihelion. \ \
(4) Aperture used for flux extraction, expressed in pixels (1 pixel =
0.25\arcsec). \ \
\end{table*}

\begin{figure}
\scalebox{0.43}{
\includegraphics*[0.5cm,6.2cm][23cm,25cm]{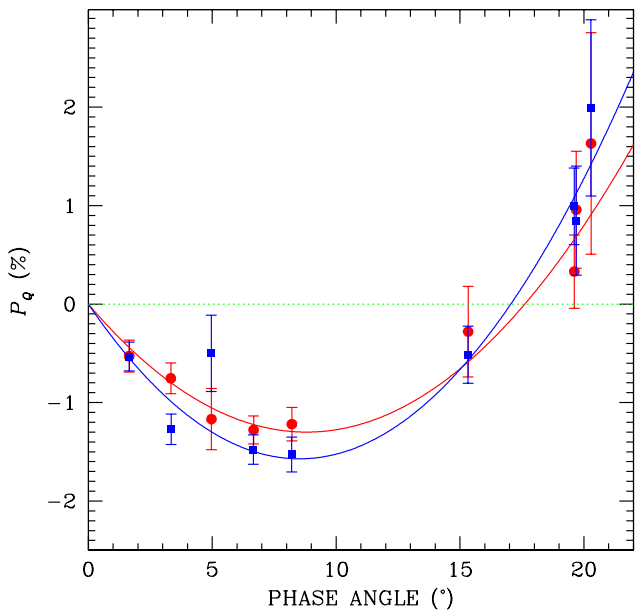}}
\caption{\label{Fig_Pq}
The measured \pq\ values as a function of the phase angle in the
$R$ (red circles) and $V$ (blue squares) filters. The solid lines
show the best-fit relation obtained with a second order polynomial
expansion constrained to pass through (0,0).
}
\end{figure}

Polarimetry was calculated using the method outlined, e.g., by
\citet{Bagetal06} and \citet{Bagetal08}. Null parameters
\citep[see][]{Bagetal09} were also systematically calculated to check
the reliability of the results.
The final reduced Stokes parameters \pq\ and \pu\ were obtained by
adopting as a reference direction the perpendicular to the great circle
passing through the object and the Sun, using Eq.~(5) of
\citet{Bagetal06}. In this way, $\pq$ represents the flux perpendicular
to the plane defined by the Sun, Object, and Earth (the scattering plane) minus the flux
parallel to that plane, divided by the sum of these
fluxes. Our final polarimetric measurements are reported in
Table~\ref{Tab_ElPiza}. Figure~\ref{Fig_Pq} shows \pq\ results as
a function of the phase angle in both filters. 

\begin{figure}
\scalebox{0.45}{
\includegraphics*[1cm,5.7cm][24cm,25cm]{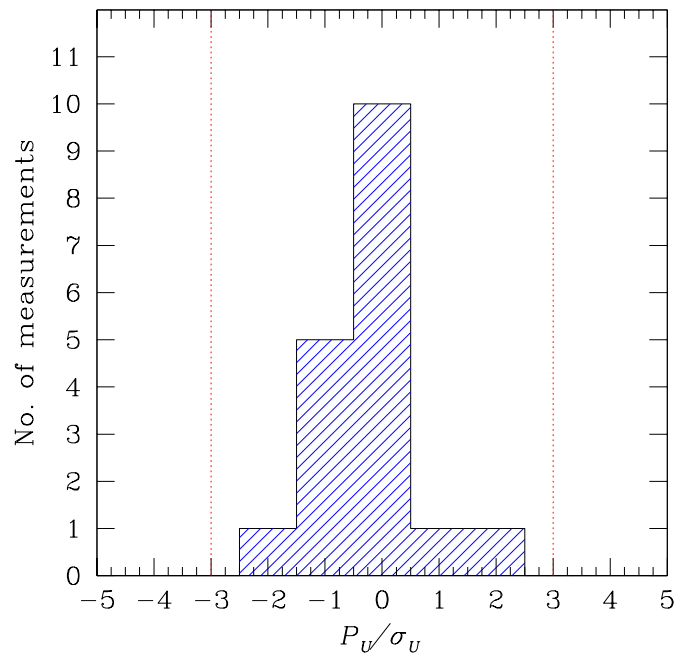}}
\caption{\label{Fig_Histogram}
Distribution of the \pu\ values expressed in units of errorbar. 
The vertical dotted lines refer to the 3-$\sigma$ detection level.}
\end{figure}

For symmetry reasons, $\pu$ values are always expected to be zero, and
inspection of their values allows us to perform an indirect quality
check of the \pq\ values. Figure~\ref{Fig_Histogram} shows the
distribution of the \pu\ values in the $R$ and $V$ filters
expressed in errorbar units. Since the distribution appears to be peaked
at about zero, and none of the \pu\ values exceed a 3-$\sigma$ detection,
this test represents a positive quality check of our polarimetric
measurements of \centotrentatre.

\subsubsection{Polarimetry of the tail}\label{Sect_Tail_Polarization}
We attempted to measure the polarization of the tail, but, not
unexpectedly, the low signal-to-noise ratio prevented us from obtaining
accurate measurements.  The highest precision was reached in the
images obtained on 13 July 2007 and 17 July 2007, where \pq\ and \pu\
could be measured with a $\sim 3$\,\% errorbar. In the $R$ filter, we
integrated the signals over an area of 110 and 130 pixels, respectively,
and we obtained $\pq=-1.60\pm 3.0$\,\% and $\pu=-0.5\pm 3.0$\,\% on
13 July 2007, and $\pq=-1.1\pm 2.5$\,\% and $\pu=0.4\pm 2.5$\,\% on
17 July 2007. On the image obtained on 5 August 2007, we integrated over a
40 pixel area of the tail, and we obtained $\pq=-1.0\pm 3.8$\,\%
$\pu=1.6\pm 3.8$\,\%.  In all cases, the null parameters were
consistent with zero at the 1.5\,$\sigma$ level.

\subsubsection{Identifying the polarization of the nucleus}
We now discuss whether our measurements can be considered
representative of the nucleus of \centotrentatre, and to what extent
they are contaminated by the coma and tail of the object.

We used the results obtained in the polar coordinate system to
estimate the true fluxed produced by the nucleus of \centotrentatre\
within the aperture used for our polarimetric measurements (see col.~8
of Table~\ref{Tab_ElPiza}), and we found that for the apertures
selected for the polarimetric measurements, the contributions from
the coma were between 0.05 and 0.1\,mag, while the contribution
from tail was generally 0.02--0.03\,mag. We can therefore assume that,
for the aperture values used for polarimetry, at least about 90\,\% of
the measured flux is related to the nucleus contribution. Assuming
  that the flux contribution due to coma and tail is entirely produced
  by a dust scattering mechanism rather than gas emission, we conclude
  that the radiation due
  to coma and tails is either parallel to the light reflected by the
  nucleus, or perpendicular to it. This hypothesis is 
  supported by all of our \pu\ measurements being consistent
  with zero. Following Eq.~(1) of \citet{Bagetal08}, we can thus
  write the nucleus polarization \pn\ as
\begin{equation}
\pn = 1.1\,\pobs - 0.1\,\pcoma \; ,
\end{equation}
where \pobs\ is the total observed polarization, and \pcoma\ is the
fraction of linear polarization of the light scattered by coma and
tail. For instance, if coma and tail are not polarized (\pcoma=0), the
nucleus polarization is underestimated (in absolute value) by about
10\,\%; if coma and tails are polarized at a 3\,\% level parallel to
the scattering plane ($\pcoma=-3$\,\%), the observed polarization minimum
of about $-1.5$\,\% would correspond, for the nucleus, to
$\sim -1.35$\,\%; if coma and tails are polarized at a 3\,\% level
perpendicular to the scattering plane (for the case of Fresnel-reflection
mechanism), the observed polarization minimum will be
$\pn=-1.95$\,\%. We note that our direct measurements presented in
Sect.~\ref{Sect_Tail_Polarization},
although affected by large errorbars, seem more consistent with a
negative value of the tail polarization than a positive one, and
infer that the fraction of tail linear
polarization is very similar to the fraction of
polarization from the nucleus plus tail.  Our dataset does not allow
us to go beyond these qualitative statements, yet our conclusion is
that the observed polarization is representative of the
polarization of the nucleus of \centotrentatre.  The following
analysis will be based on this conclusion.

\subsubsection{Characterization of the polarimetric curve}
As for the large number of atmosphere-less solar system bodies,
the nucleus of the \centotrentatre\ exhibits the phenomenon of
\textit{negative} polarization: at small phase angles, the electric
field vector component parallel to the scattering plane predominates
over the perpendicular component, in contrast to what is expected from
the simple single Rayleigh-scattering or Fresnel-reflection
model. This phenomenon is generally explained in terms of coherent
backscattering \citep[e.g.,][]{Muinonen04}. For small bodies of the
solar system, the polarization reaches a minimum value generally
between phase angle 7\degr and 10\degr, and becomes positive at phase
angles of between 17\degr\ and 22\degr\ \citep[e.g.,][]{Penetal05}.  In conclusion,
for those small bodies for which it is possible to measure the
behaviour of the polarization for an extended range of phase angles,
three important characteristics can be measured well, and used to perform an
empirical classification: the minimum polarization and its corresponding phase
angle, the slope of the polarimetric curve in the linear part
beyond polarization minimum, and the inversion angle at which the
polarization changes from being parallel to the scattering plane and
becomes, at larger phase angles, perpendicular to the scattering
plane. Following these guidelines, we perform an analysis of our
polarimetric data for the nucleus of 133/Elst-Pizarro.

Inspection of Fig.~\ref{Fig_Pq} shows that measurements in both the $R$ and
$V$ filters agree within the errors (except for a
point at phase angle 3\degr). Nevertheless, we perform our basic
analysis based on the measurements obtained in the $R$ and $V$
filters separately.

The best-fit relation obtained using a second order polynomial (constrained to pass
through the origin) shows that in the $R$ filter, the minimum is reached
at phase angle $\alpha_R^\mathrm{min}=8.8\degr \pm 0.6\degr$, for a
minimum polarization value of $\pq^\mathrm{min}(R)=-1.30 \pm
0.20$\,\%. In the $V$ filter, the minimum is reached at phase angle
$\alpha_V^\mathrm{min}=8.5\degr \pm 0.5\degr$, with a minimum
polarization value of $\pq^\mathrm{min}(V)=-1.57 \pm 0.20$\,\% .
The positive crossover, when the observed polarization changes sign from
 negative to positive, is at phase angle $\alpha_R(\pq=0) = 17.6\degr \pm
2.1\degr$ in the $R$ filter, and at phase angle $\alpha_V(\pq=0) = 17.0\degr
\pm 1.6\degr$ in the $V$ filter.

Using a second order polynomial, the slope of the polarization depends
clearly on the phase angle in which it is calculated. With reference to the
the crossover point, the polarization slopes are
\[
\left(\frac{\mathrm{d}\pq(R)}{\mathrm{d}\alpha}\right)_{\pq=0} = 
0.30 \pm 0.02\,\%{\rm /deg}
\]
and
\[
\left(\frac{\mathrm{d}\pq(V)}{\mathrm{d}\alpha}\right)_{\pq=0} = 
0.37 \pm 0.02\,\%{\rm /deg} \, ,
\]
for the $R$ and the $V$ filter, respectively. We note however that these
estimates are obtained based on the assumption that the
measurements may be fitted with a second order polynomial.

\subsubsection{Relationships between polarization and albedo}
For asteroids, two empirical relationships have been found between the
polarization characteristics and the albedo, namely, both slope and
polarization minimum seem to be related to the geometric albedo of the body 
by means of a simple expression as given, e.g., by \citet{LupMoh96}
\begin{equation}
\log(p) = -0.98 
\log\left(\frac{\mathrm{d}\pq}{\mathrm{d}\alpha}\right)_{\pq=0} - 1.73 \label{Eq_Slope}
\end{equation}
and
\begin{equation}
\log(p) = -1.22 
\log\left(\pq^\mathrm{min}\right) - 0.92                     \label{Eq_Minimum} \; ,
\end{equation}
where $p$ is the albedo, $\mathrm{d}\pq/\mathrm{d}\alpha$ is the
derivative of the observed linear polarization with respect to phase
angle, and $\pq^\mathrm{min}$ is the minimum value of the
polarization. Using Eq.~(\ref{Eq_Slope}), from our data we obtain for
the albedo the values $p_R = 0.061 \pm 0.004$ in the $R$ filter, and
$p_V = 0.049 \pm 0.003$ in the $V$ filter.  Using
Eq.~(\ref{Eq_Minimum}), we obtain $p_R = 0.087 \pm 0.017$ and $p_V =
0.069\pm 0.011$. Their average values correspond to about $p_R= 0.07$
and $p_V= 0.06$, uncertainties being about 0.01.

The average albedo value obtained from our polarization measurements
agrees remarkably well with that determined via the classical
method of combined visible and thermal flux measurements of
\centotrentatre, i.e., $p_R=0.05\pm0.02$ as published by
\citet{Hsietal09}.  However, it remains to be shown that the
albedo-polarization relationships are indeed applicable to our target
since they were verified and calibrated using asteroid data for
different taxonomic types than those that we have found for
\centotrentatre\ (see Sect.~\ref{Sect_Discussion}).

\section{Discussion}\label{Sect_Discussion}
There are two main aspects to the properties of \centotrentatre. On
the one hand, it is located in the asteroid belt with an orbit similar
to the Themis collision family or a subgroup of it, the Beagle family,
and over a large part of its orbital revolution it appears
asteroid-like. On the other hand, around and after perihelion it
displays activity producing a dust tail, as comets do. In the
following, we discuss whether our new polarimetric measurements could help us to
identify whether it is an asteroid (with cometary activity) or a comet
(scattered into the asteroid belt).
\begin{figure*}
\scalebox{0.90}{
\includegraphics*[1.2cm,6.4cm][22cm,16cm]{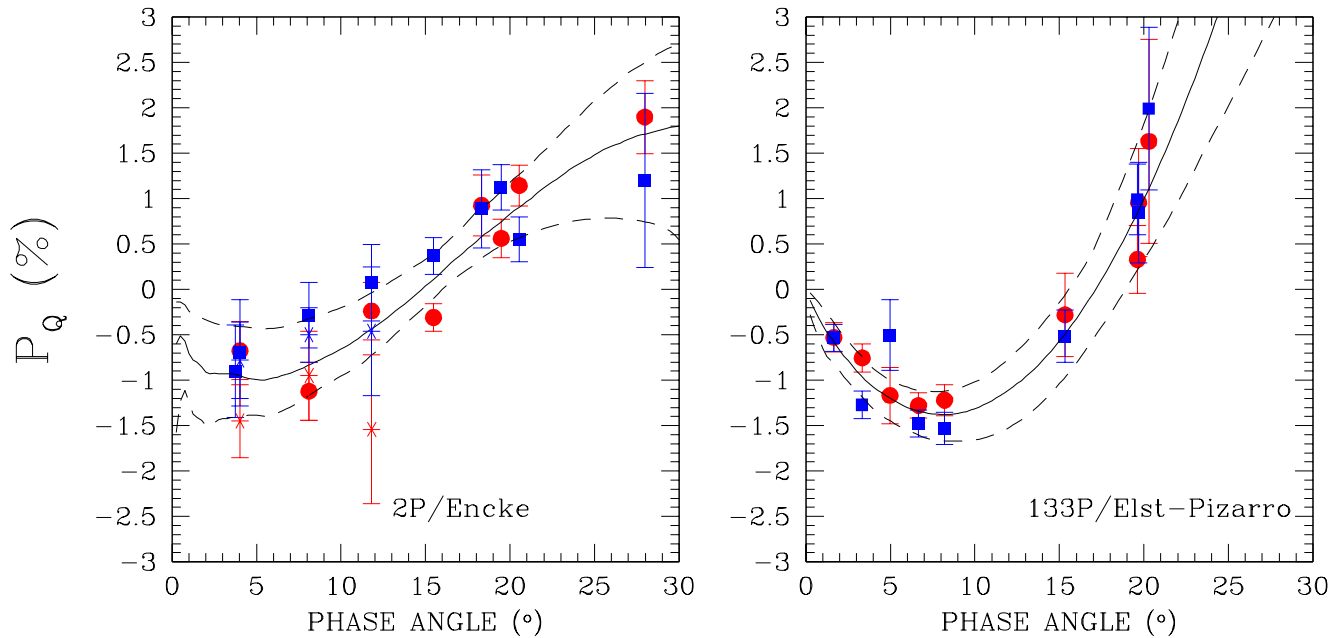}}
\caption{\label{Fig_MCMC} 
   Polarimetric observations of the nucleus of comet 2P/Encke (left panel)
   and of \protect\centotrentatre\ (right panel), modeled using MCMC
   for the empirical linear-exponential model. Blue squares refer to
   $V$ Bessell filter, and red circles to $R$ Bessell filter. For 2P/Encke,
   red and blue asterisks refer to narrow band filters centred about
   834 and 485\,nm, respectively. The best-fit models are
   shown with solid line, while the 3\,$\sigma$ error envelopes are
   shown with dashed lines. 
}
\end{figure*}

We first compare the polarimetric curve of the nucleus of
\centotrentatre\ with that of the nucleus of comet 2P/Encke
\citep{Boeetal08} by making use of the scattering parametrisation
provided in \citet{Muietal09a,Muietal09b}. By using lines of electric
dipoles with an inverse-gamma distribution for the line lengths, and
by estimating the parameters with a Markov-Chain Monte-Carlo (MCMC)
method \citep[see, e.g.,][]{Muietal09b}, we have computed, for
133P/Elst-Pizarro and 2P/Encke, physical interference models. The
eight parameters of the full model are: the exponents or shape
parameters $s_1$ and $s_2$ and the scale parameters $y_1$ and $y_2$ of
the inverse-gamma distributions for the two polarization states
(subscripts 1 and 2 are for positive and negative states,
respectively); the complex amplitude $Z$ and interdipole distance
$kd_3$ ($k=2\pi/\lambda$ is the wave number and $d_3$ the physical
distance) for the longitudinal electric-dipole contribution; and the
normalized weight $w$ of the Rayleigh-like contribution to the
polarization curve.

For 133P/Elst-Pizarro and 2P/Encke, reduced six parameter models already
suffice to explain the polarimetric
observation. We assume that $s_1=s_2=s$
and fix $kd_3=\pi$. Least squares fitting analysis and MCMC sampling provide the
polarization curve corresponding to the best-fit relations (solid lines), and in the
$3\,\sigma$ error envelopes (dashed lines) depicted in Fig.~\ref{Fig_MCMC}.
Best-fit model parameters and their $3\,\sigma$ errors are given in Table~\ref{Tab_MCMC}.
\begin{table}
\caption{\label{Tab_MCMC} MCMC parameters for \protect\centotrentatre, 2P/Encke, and
(1) Ceres. $\Re(Z)$ and $\Im(Z)$ are the real and imaginary parts of the complex
amplitude $Z$, respectively.}
\begin{tabular}{llll}
\hline\hline
         &\multicolumn{1}{c}{133P/Elst-Pizarro}&\multicolumn{1}{c}{2P/Encke}&\multicolumn{1}{c}{(1) Ceres}\\
\hline
$s$      & $\phantom{-}0.87^{+0.47}_{-0.31}$&$0.31^{+0.27}_{-0.24}$   &  $0.731^{+0.051}_{-0.024}$\\[2mm]
$y_1$    & $\phantom{-}2.39^{+2.19}_{-0.84}$   & $0.99^{+0.41}_{-0.26}$   &  $3.258^{+0.281}_{-0.079}$\\[2mm]
$y_2$    & $\phantom{-}0.412^{+0.033}_{-0.412}$ & $0.236^{+0.064}_{-0.047}$& $0.2053^{+0.0023}_{-0.0159}$\\[2mm]   
$\Re(Z)$ & $-0.20^{+0.56}_{-0.50}$  & $1.15^{+0.31}_{-0.22}$   & $0.0037^{+0.0385}_{-0.1000}$    \\[2mm]
$\Im(Z)$ & $-0.39^{+1.82}_{-0.36}$  & $1.975^{+0.328}_{-0.099}$ & $2.0607^{+0.0996}_{-0.0048}$   \\[2mm]
$w$      & $\phantom{-}0.71^{+0.15}_{-0.21}$   & $0.848^{+0.36}_{-0.045}$  &$0.8084^{+0.0070}_{-0.0096}$\\[1mm]
\hline
\end{tabular}
\end{table}
Among the parameters, the weights $w$, exponents $s$ and scale
parameters $y_1$ and $y_2$ are of importance for the present study.
The weights $w$ obtained realistic values $w > 0.5$, which, in the
physical model, indicate a positive contribution from point-like
Rayleigh scatterers in addition to the contributions from the lines of
dipoles. The exponents $s$ and the scale parameters $y_1$ and $y_2$
are loosely constrained by the polarimetric data in the current case
of considerable observational errors.

In addition to 2P/Encke and \centotrentatre, the physical model has so
far been applied only to the polarimetric data of asteroid (1)~Ceres
\citep{Muietal09b}. The corresponding best-fit model parameters are given
in Table~\ref{Tab_MCMC}.

We conclude that it is too early to reach definitive conclusions using
MCMC models, beyond that the polarization characteristics of
133P/Elst-Pizarro are closer to those of (1)~Ceres than those of
2P/Encke. The polarimetric slope at the inversion angle is larger for
133P/Elst-Pizarro than for (1)~Ceres. All three weights $w$ can
overlap within their 3-$\sigma$ error domains and the same is true for
the scale parameter $y_2$. Mainly in terms of the scale parameter
$y_1$, 2P/Encke differs from the others.

A more statistically meaningful approach is offered by a straight
comparison of the photometric and polarimetric properties of
\centotrentatre\ with those of different types of asteroids, cometary
dust, Centaurs and Kuiper-Belt objects. Table~2 of \citet{Boeetal08}
lists the typical numerical values for albedo, slope of the
photometric phase function, spectral gradient, minimum polarization
and the related phase angle, slope of the polarization phase function,
inversion angle, and spectral gradient of the polarization for
different classes of small bodies of the solar system. A comparison
with the values of \centotrentatre\ derived in this work allows us to
conclude that \centotrentatre\ does not exhibit the typical
properties of Kuiper-Belt objects and Centaurs, C, S, E and M
type asteroids, cometary dust, while there is a rather good
agreement in all parameters with those of F-type asteroids. 

F-type asteroids are considered to be primitive \citep{Busetal02}, and
they are sometimes claimed to be related to comet-like objects
\citep{Weietal02}. A weakness in this comparison is the meager
knowledge of polarimetric parameters of certain types of small bodies,
such as cometary nuclei, Centaurs, and also F-type asteroids themselves
(although the polarimetric dataset for F-type asteroids is
substantially larger than that for Centaurs and cometary
nuclei). However, if the parameter similarity between
\centotrentatre\ and F-type asteroids is to be seen as an indicator
of the similarity of these objects, one can predict that more objects
with cometary activity will be found among asteroids with F-type
taxonomy.

\citet{Licetal07a} assigned \centotrentatre\ a C- or B-type classification
taxonomy classes to which F-type asteroids
are closely related to (the water absorption feature around 3$\mu$m is
missing and differences in the continuum below 0.4$\,\mu$m exist for
F-type objects). Other notable, peculiar C-class objects include
107P/Wilson-Harrington \citep[C- or F-type;][]{Tholen1989}, 3200
Phaethon \citep[B- or F-type;][]{Tholen1989,Licetal07b}, and 2005 UD
\citep[B- or F-type;][]{Jewitt06,Kinetal07}. The former two
objects at least have displayed cometary activity in the past. An F-type
\centotrentatre\ would thus have good company with respect to cometary
activity.

Comet migration into the asteroid belt does not seem to be a very
efficient process dynamically \citep{Levetal06}, and the ``snowline'',
i.e., the distance to the Sun at which water vapour in the
protoplanetary disk condenses and becomes accreted in forming
planetesimal, falls into the (outer) asteroid belt. Therefore, one may
be inclined to accept \centotrentatre\ as a ``child of the inner
solar system'', and not of the cold outskirts where cometary nuclei
originate. Our observational results are certainly compatible
with this scenario, although it implies that cometary activity can
also occur in asteroids. The dust activity in \centotrentatre\ and the
other so-called 'main-belt comets' require a driving mechanism. Ice
sublimation is considered a possibility, but is not as yet supported
by observational evidence.

\section{Conclusions}
With the FORS1 instrument of the 8\,m ESO VLT, we have carried out
imaging observations in polarimetric mode of the object
\centotrentatre\ at nine observing epochs. The first observation was performed
in May 2007, about one month before perihelion, and the remaining eight
from July to September 2007, up to about three months after
perihelion. These observations cover the phase-angle range of
0\degr-20\degr.

Our images have detected one or two tails close to
perihelion, and, to a marginal level of evidence, the presence of a
coma. We have then performed an analysis of the dust production.  The
dust release of \centotrentatre\ may have started about one month
before perihelion passage and continued at least until about three
months after perihelion. The onset of dust activity in 2007 happened
earlier than concluded for the 1996 apparition of the comet by
\citet{Praetal96} and \citet{Boe1996}, although, admittedly, the
latter analyses could not really constrain dust release before
perihelion passage. However, the dust release in \centotrentatre\ is
clearly repetitive and it starts and extends around perihelion
passage.  The dust grains that dominate the optical appearance of the
tail were of micrometer size and larger, and were released at a
relatively low speed. The appearance of the tail in 2007 resembles
that seen in 1996, although the two tail patterns refer to different
emission periods with respect to perihelion passage. The 1996 tail
appearance supports the scenario of dust production that continued for
up to half a year after perihelion passage. Considering the results of
\citet{Hsietal04}, the active period would even extend to about a year
after perihelion. The dust tails seen in \centotrentatre\ during the
1995, 2002, and 2007 apparitions support the concept that the activity
of \centotrentatre\ is recurrent and that it extends over the same arc
of the orbit (from about 1 month before perihelion to at least 1 year
after perihelion). A possible yet unproven scenario, 
described in greater detail by \citet{Hsietal04} and implicitly assumed 
also in earlier publications by \citet{Praetal96} and 
\citet{Boeetal97}, is that a main region of local
activity is switched on -- possibly due to solar illumination --
shortly before perihelion and continues to be active along the orbit
over a year or so. The repetitive behaviour of the activity would
suggest a rather stable orientation of the rotation axis over the past
three apparitions of \centotrentatre. The activity level of the
nucleus is not very high and produces micron-size dust at low
speed. The true release mechanism could not be constrained from our
observations.

We have used our observations to characterize the polarimetric and
photometric behaviour of the object as a function of its phase-angle,
and we have compared it to that of the nucleus of comet 2P/Encke, and
other small bodies of the solar-system. In particular, the comparison
with the polarimetric curve of 2P/Encke was performed by adopting a
physical classification tool for the polarization phase curves, but
our results are not conclusive because current lack of a statistically
meaningful sample of objects that have been analysed in a similar
fashion. However, a direct comparison of the polarimetric and
photometric curves identified several similarities with the
observational parameters of F-type asteroids.
\begin{acknowledgements}
We thank Dr.\ H.\ Hsieh for a useful discussion at the early stages
of the work, and the anonymous referee for a very careful review
of the original manuscript.
\end{acknowledgements}


\end{document}